\documentclass[12pt]{article}
\usepackage{graphicx}
\usepackage{amsmath}
\usepackage{amsbsy}
\usepackage{amsfonts}
\usepackage{amsthm}
\begin{document}

\title{Q-stars in scalar-tensor gravitational theories in extra dimensions}

\author{Athanasios Prikas}

\date{}

\maketitle

Physics Department, National Technical University, Zografou
Campus, 157 80 Athens, Greece.\footnote{e-mail:
aprikas@central.ntua.gr}

\begin{abstract}
We present Jordan-Brans-Dicke and general scalar-tensor
gravitational theory in extra dimensions in an asymptotically flat
or anti de Sitter spacetime. We consider a special gravitating,
boson field configuration, a $q$-star, in $3$, $4$, $5$ and $6$
dimensions, within the framework of the above gravitational theory
and find that the parameters of the stable stars are a few per cent
different from the case of General Relativity.
\end{abstract}

PACS number(s): 11.27.+d, 04.40.-b


\newpage

\section{Introduction - Jordan-Brans-Dicke theory in any dimension}

Boson stars are stable configurations of a massive, complex scalar
field coupled to gravity, \cite{boson-stars-old1,boson-stars-old2}.
Self interactions, \cite{boson-stars-new,boson-stars-quartic}, the
coupling of the matter scalar field to a gauge field,
\cite{boson-stars-charged}, or the rotation of the solitonic object,
\cite{boson-stars-rotating}, were taken into account. When the
non-relativistic theory admits non-topological soliton solutions,
the corresponding relativistic generalizations, which are very large
field configurations, are the soliton stars,
\cite{large-soliton-stars1,large-soliton-stars2,large-soliton-stars3}.
Q-balls is a special class of non-topological solitons, appearing in
Lagrangians with a global $U(1)$ symmetry, \cite{qballs-initial}, a
local $U(1)$, \cite{qballs-charged}, or a non-abelian $SU(3)$ or
$SO(3)$ symmetry, \cite{qballs-nonabelian}. Q-stars have been
investigated in various models, with one and two scalar fields,
\cite{qstars-global}, non-abelian symmetries,
\cite{qstars-nonabelian}, with a scalar and a fermion field,
\cite{qstars-fermion} and with a local $U(1)$ symmetry,
\cite{qstars-charged}, in asymptotically anti de Sitter spacetime,
\cite{qstars-AdS}, and in 4 or more dimensions,
\cite{qstars-extradim}.

Scalar-tensor gravitational theories appeared in the original papers
of Brans and Dicke, and Jordan, \cite{bransdicke,jordan}. The
Newtonian constant $G$ is replaced by the inverse mean value of a
scalar field, $\phi_{\textrm{BD}}$, and the total action contains
kinetic terms for the new field times an $\omega_{\textrm{BD}}$
quantity, which is regarded as a constant in the original theory.
The theory can be generalized by replacing the constant
$\omega_{\textrm{BD}}$ by a function, usually of the Brans-Dicke
(BD) scalar field, \cite{scalartensor1,scalartensor2}. The theory of
a scalar field, with quartic self-interactions, coupled to the
metric and the BD scalar formulated in \cite{bosonstar-bransdicke1}.
Various properties of such configurations have been studied in a
series of papers,
\cite{bosonstar-bransdicke2,bosonstar-bransdicke3}, where especially
the matter of gravitational memory of boson stars in scalar tensor
gravity has been analyzed,
\cite{bosonstar-bransdicke4,bosonstar-bransdicke5}. Their results
generalized in scalar-tensor gravitational theories with
$\omega_{\textrm{BD}}$ no more a constant, but a function of the BD
scalar,
\cite{bosonstar-scalartensor1,bosonstar-scalartensor2,bosonstar-scalartensor-charged}.

The purpose of the present work is double: We write the BD
gravitational theory in $D$ dimensions and apply our results in a
realistic case of a scalar field, admitting q-ball type solutions
in the absence of gravity. We compare our results with the
corresponding ones obtained in General Relativity and we
investigate the influence of the spacetime dimensionality in the
parameters of the star.

All the field configurations under consideration are spherically
symmetric and static, so we use a static, spherically symmetric
metric in $D$ dimensions:
\begin{equation}\label{1}
ds^2=-e^{\nu}dt^2+e^{\lambda}d\rho^2+\rho^2d\Omega_{D-2}^2\ ,
\end{equation}
with $g_{tt}=-e^{\nu}$ and $d\Omega_{D-2}^2$ the line element on
a $(D-2)-$dimensional unit sphere. The action is:
\begin{equation}\label{2}
S=\frac{1}{16\pi}\int
d^Dx\sqrt{-g_D}\left[\phi_{\textrm{BD}}(R-2\Lambda)
-\frac{\omega_{\textrm{BD}}g^{\mu\nu}
\partial^{\lambda}\phi_{\textrm{BD}}\partial_{\lambda}\phi_{\textrm{BD}}}{\phi_{\textrm{BD}}^2}
\right]+S_{\textrm{matter}}\ ,
\end{equation}
with $S_{\textrm{matter}}$ the contribution to the total action
of the matter fields. $\Lambda$ stands for the negative, or zero,
cosmological constant. The Einstein equations take the form:
\begin{align}\label{3}
G_{\mu\nu}=\frac{8\pi}{\phi_{\textrm{BD}}}T_{\mu\nu}+\frac{\omega_{\textrm{BD}}}
{\phi_{\textrm{BD}}^2}\left(\partial_{\mu}\phi_{\textrm{BD}}\partial_{\nu}\phi_{\textrm{BD}}-
\frac{1}{2}g_{\mu\nu}\partial^{\lambda}\phi_{\textrm{BD}}\partial_{\lambda}\phi_{\textrm{BD}}
\right)+\nonumber \\ \frac{1}{\phi_{\textrm{BD}}}
\left(\phi_{\textrm{BD},\mu;\nu}-g_{\mu\nu}
{\phi_{\textrm{BD}}}_{;\lambda}^{\hspace{1em};\lambda}\right)-\Lambda
g_{\mu\nu} \ ,
\end{align}
where $T_{\mu}^{\ \nu}=\textrm{diag}(-\varepsilon,p,...p)$ is the
energy momentum tensor with trace $T=\varepsilon+Dp$. The
equation of motion for the BD field is:
\begin{equation}\label{4}
\frac{2\omega_{\textrm{BD}}}{\phi_{\textrm{BD}}}
{\phi_{\textrm{BD}}}_{;\lambda}^{\hspace{1em};\lambda}-\frac{\omega_{\textrm{BD}}
\partial^{\lambda}\phi_{\textrm{BD}}\partial_{\lambda}\phi_{\textrm{BD}}}{\phi_{\textrm{BD}}^2}
+R-2\Lambda=0\ .
\end{equation}
Contracting eq. \ref{3}, and substituting the trace in eq. \ref{4}
we find the final form for the equation of motion of the BD field:
\begin{equation}\label{5}
{\phi_{\textrm{BD}}}_{;\lambda}^{\hspace{1em};\lambda}=\frac{8\pi
T-2\Lambda\phi_{\textrm{BD}}}{(D-2)\omega_{\textrm{BD}} +D-1}\ .
\end{equation}
Equation \ref{5} reveals the dimensional sensitivity of the BD
theory. Both $T$ and the $\omega_{\textrm{BD}}$-dependent factor
depend on the spacetime dimensionality.

We will now find the asymptotic relation for the BD field. For
$p\ll\varepsilon$ and $\Lambda=0$, the $tt$ component of the
Einstein equations gives:
\begin{equation}\label{6}
G_t^{\
t}=-\frac{8\pi\varepsilon}{\phi_{\textrm{BD}}}\frac{(D-2)\omega_{\textrm{BD}}+D}
{(D-2)\omega_{\textrm{BD}}+D-1}\ .
\end{equation}
The corresponding result of the Einstein gravity is:
\begin{equation}\label{7}
G_t^{\ t}=-8\pi G_D\varepsilon\ .
\end{equation}
We know that for $\varepsilon\ll$, the limit of the BD theory is
the Einstein gravity. So, for localized matter configurations,
the right boundary value at infinity for the BD field is:
\begin{equation}\label{8}
\phi_{\textrm{BD}}=\frac{1}{G_D}
\frac{(D-2)\omega_{\textrm{BD}}+D}{(D-2)\omega_{\textrm{BD}}+D-1}
\ .
\end{equation}
One can verify that eqs. \ref{6} and \ref{8} give the right result
for the well-known $4$-dimensional case. We will now apply our
results to a certain case of gravitating scalar matter, namely
q-stars.

\section{Q-stars in BD gravitational theory}

\begin{figure}
\centering
\includegraphics{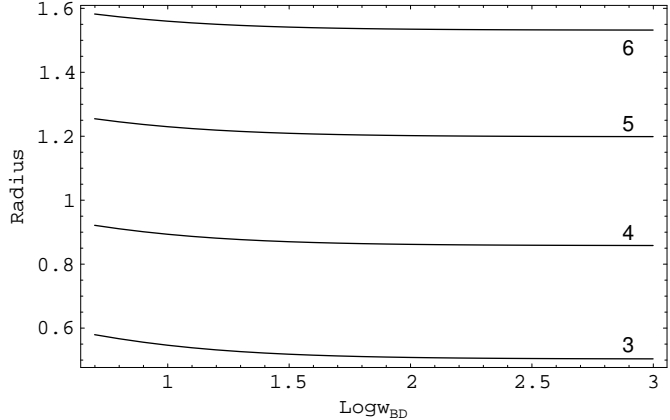}
\caption{The radius of a q-star as a function of
$\omega_{\textrm{BD}}$. The numbers within the figures denote the
spacetime dimensionality. For $\omega_{\textrm{BD}}\simeq1000$
the results of General Relativity are approximately reproduced. In
figures \ref{figure1}-\ref{figure4} and
\ref{figure6}-\ref{figure14} we use $\omega=0.45$, equivalently
$A_{\textrm{sur}}=0.81$. When decreasing $\omega_{\textrm{BD}}$,
the star parameters are a few per cent larger.} \label{figure1}
\end{figure}

\begin{figure}
\centering
\includegraphics{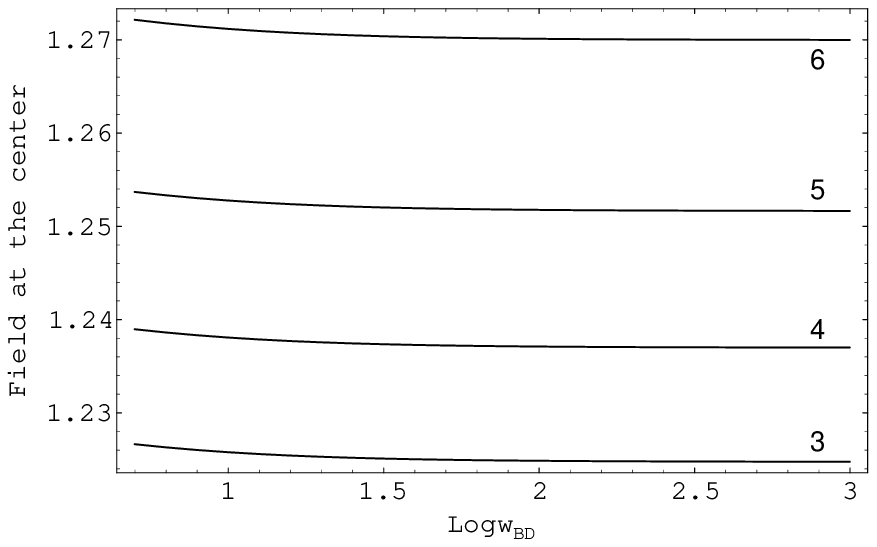}
\caption{The value of the matter scalar field at the center of
the q-star as a function of $\omega_{\textrm{BD}}$.}
\label{figure2}
\end{figure}

\begin{figure}
\centering
\includegraphics{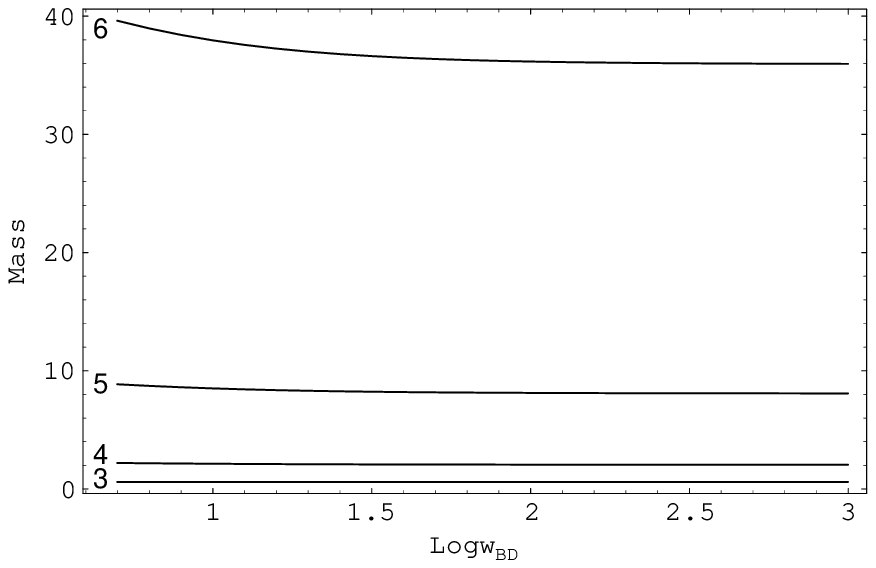}
\caption{The total energy of a q-star as a function of
$\omega_{\textrm{BD}}$.} \label{figure3}
\end{figure}

\begin{figure}
\centering
\includegraphics{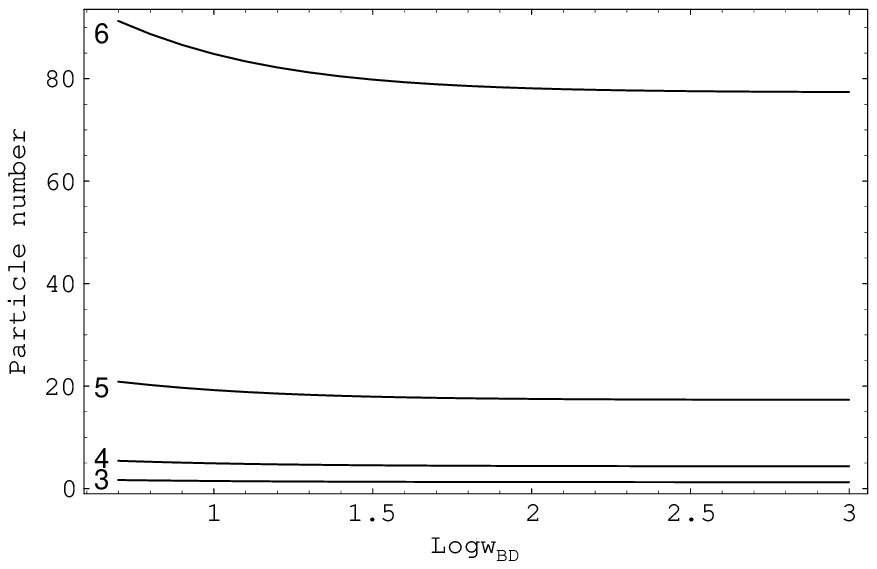}
\caption{The particle number of a q-star as a function of
$\omega_{\textrm{BD}}$.} \label{figure4}
\end{figure}

\begin{figure}
\centering
\includegraphics{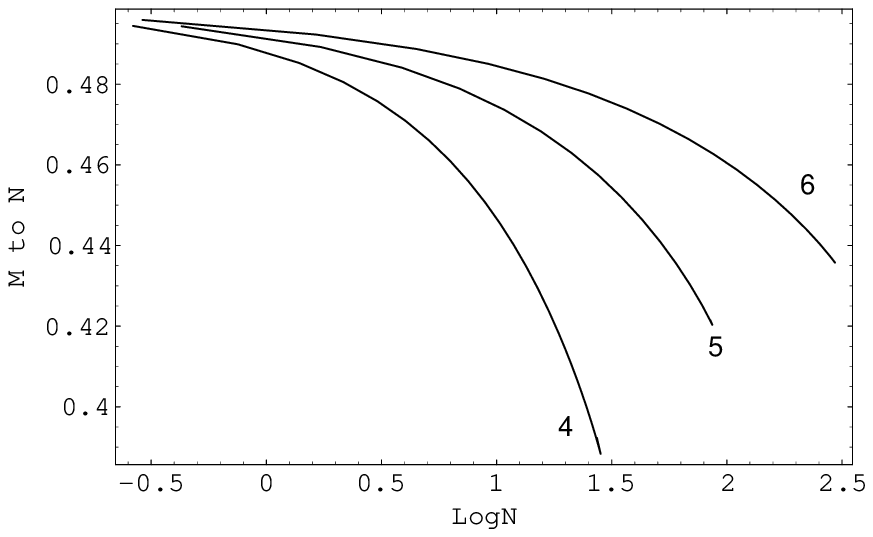}
\caption{The ratio of the star mass to its particle number as a
function of the particle number for $\omega_{\textrm{BD}}=500$.
Small values of the particle number correspond to small stars, whose
limiting value is the non-gravitating q-ball.} \label{figure5}
\end{figure}

The Lagrangian resulting from the presence of a matter scalar
field is:
\begin{equation}\label{9}
\mathcal{L}_{\textrm{matter}}=-(\partial_{\mu}\phi)^{\ast}(\partial^{\nu}\phi)-U\
,
\end{equation}
where $\phi$ is the matter scalar and $U$ a suitable potential,
admitting q-ball type solutions in the absence of gravity. We will
now insert the q-soliton ansatz:
\begin{equation}\label{10}
\phi(\vec{\rho},t)=\sigma(\rho)e^{-\imath\omega t}\ .
\end{equation}
The equation of motion for the scalar field:
\begin{equation}\label{11}
\phi_{;\lambda}^{\hspace{1em};\lambda}-\frac{dU}{d|\phi|^2}\phi=0
\ ,
\end{equation}
now takes the form:
\begin{equation}\label{12}
\sigma''+\left[\frac{D-2}{\rho}+\frac{1}{2}(\nu'-\lambda')\right]\sigma'+
e^{\lambda}\omega^2e^{-\nu}\sigma-e^{\lambda}\frac{dU}{d\sigma^2}\sigma=0\
,
\end{equation}
where the prime denotes the derivative with respect to $\rho$. We
define:
\begin{equation}\label{13} A=e^{-\lambda}\ ,
\hspace{1em} B=e^{-\nu}\ ,
\end{equation}
\begin{equation}\label{14}
\begin{split}
W&\equiv e^{-\nu}{\left(\frac{\partial\phi}{\partial
t}\right)}^{\ast}\left(\frac{\partial\phi}{\partial t}\right)=
e^{-\nu}{\omega}^2{\sigma}^2\ , \\ V&\equiv
e^{-\lambda}{\left(\frac{\partial\phi}{\partial\rho}\right)}^{\ast}
\left(\frac{\partial\phi}{\partial\rho}\right)=
e^{-\lambda}{\sigma'}^2
\end{split}
\end{equation}
and rescale:
\begin{equation}\label{15}
\begin{split}
\tilde{\rho}=\rho m\ , \hspace{1em} \tilde{\omega}&=\omega/m\ ,
\hspace{1em} \tilde{\phi}=\phi/m^{\frac{D-2}{2}} \ , \\
\widetilde{U}=U/m^D\ , \hspace{1em} \widetilde{W}&=W/m^D\ ,
\hspace{1em} \widetilde{V}=V/m^D\ .
\end{split}
\end{equation}
We also rescale for convenience the BD scalar field:
\begin{equation}\label{16}
\Phi_{\textrm{BD}}=\phi_{\textrm{BD}}G_D\frac{(D-2)\omega_{\textrm{BD}}+D-1}
{(D-2)\omega_{\textrm{BD}}+D}\ .
\end{equation}
Comparing eqs. \ref{8}, \ref{16} we find that
$\Phi_{\textrm{BD}}\rightarrow1$ for $\rho\rightarrow\infty$ and
for localized field configurations.

Gravity becomes important when $R^{D-3}\sim G_DM(R)$ where $M(R)$
is a parameter, dependent on the mass trapped within a sphere of
radius $R$. We define:
\begin{equation}\label{17}
\epsilon\equiv\sqrt{8\pi G_Dm^{D-2}}\ ,
\end{equation}
which is a very small quantity for $m$ of the order of magnitude
of some (hundreds) $GeV$. We find for the interior of a q-star
that $U\sim W\sim m^D$, $V\sim\epsilon^2m^D$, so the energy
resulting from the spatial variation of matter field can be
neglected within the star. We also define:
\begin{equation}\label{18}
\tilde{r}=\epsilon\tilde{\rho}\ .
\end{equation}
We use a rescaled potential:
\begin{equation}\label{19}
\widetilde{U}=|\tilde{\phi}|^2\left(1-|\tilde{\phi}|^2+\frac{1}{3}|\tilde{\phi}|^4\right)
=\tilde{\sigma}^2\left(1-\tilde{\sigma}^2+\frac{1}{3}\tilde{\sigma}^4\right)\
.
\end{equation}
This potential admits q-ball type solutions in the absence of
gravity with
$1>\omega\geq({\sqrt{U/|\phi|^2}})_{\textrm{min}}=1/2$. Dropping
from now on the tildes and the $O(\epsilon)$ quantities, we find
an analytical solution for the scalar field:
\begin{equation}\label{20}
\sigma=(1+\omega B^{1/2})^{1/2}\
,\hspace{1em}W=B\omega^2(1+\omega B^{1/2})\
,\hspace{1em}U=\frac{1}{3}(1+\omega^3B^{3/2})\ .
\end{equation}

The eigenvalue equation for the frequency $\omega$ of the scalar
field can be found from the equation of motion of the matter field
within the surface. As we know from
\cite{qstars-global,qstars-nonabelian,qstars-fermion}, q-type
solitons are characterized by a large interior where the scalar
field is approximately constant and a thin surface of width of order
of $m^{-1}$, where the scalar field varies rapidly from the $\sigma$
value at the inner edge of the surface, to the zero value at the
outer edge. The soliton exterior is clearly distinguished from the
soliton interior and the surface, because the scalar field vanishes
at the exterior. The situation is different in the case of
\emph{non-solitonic} boson stars, where the star radius is of order
of $m^{-1}$, the scalar field varies from $\sim M_{\textrm{Pl}}$ at
the center of the soliton to the zero value outside and there is no
clear distinction between the star interior and the star surface. In
the case of solitonic stars, the soliton radius is uniquely defined
by the outer edge of the surface. Dropping the $O(\epsilon)$
quantities form the Lagrange equation within the surface we find:
\begin{equation}\label{21}
V+W-U=0\ .
\end{equation}
The above equation holds true only within the surface. At the
inner edge of the surface $\sigma'$ and, consequently, $V$ are
zero, in order to match the interior with the surface solution,
so $W=U$. Using also eq. \ref{20} we find:
\begin{equation}\label{22}
\omega=\frac{A_{\textrm{sur}}^{1/2}}{2}=\frac{B_{\textrm{sur}}^{-1/2}}{2}\
,
\end{equation}
where $A_{\textrm{sur}}$, $B_{\textrm{sur}}$ denote the value of
the metrics at the surface of the star. In the absence of
gravity, $A(r)=B(r)=1$, so $\omega=1/2$.

We will now turn to the Einstein equations. In $D$ dimensions,
\begin{equation}\label{23}
\begin{split}
G_t^{\ t}&=\frac{A-1}{2r^2}(D-3)(D-2)+\frac{A'}{2r}(D-2)\ ,\\
G_r^{\ r}&=\frac{A-1}{2r^2}(D-3)(D-2)-\frac{AB'}{2Br}(D-2)\ .
\end{split}
\end{equation}
The energy-momentum tensor for the matter scalar field is:
\begin{equation}\label{24}
T_{\mu\nu}={({\partial}_{\mu}\phi)}^{\ast}({\partial}_{\nu}\phi)+
({\partial}_{\mu}\phi){({\partial}_{\nu}\phi)}^{\ast}
-g_{\mu\nu}[g^{\alpha\beta}{({\partial}_{\alpha}\phi)}^{\ast}({\partial}_{\beta}\phi)]
-g_{\mu\nu}U\ .
\end{equation}
Using eqs. \ref{10}, \ref{13}-\ref{15} and \ref{20} and dropping
the $O(\epsilon)$ quantities we can write the Einstein equations:
\begin{align}\label{25}
G_t^{\
t}=&\frac{(D-2)\omega_{\textrm{BD}}+D-1}{[(D-2)\omega_{\textrm{BD}}+D]
\Phi_{\textrm{BD}}}\times \nonumber\\ &\left[-W-U-\frac{(D-2)W-DU
-2\Lambda\Phi_{\textrm{BD}}\frac{(D-2)\omega_{\textrm{BD}}+D}
{(D-2)\omega_{\textrm{BD}}+D-1}}{(D-2)\omega_{\textrm{BD}}+D-1}\right]
\nonumber\\
&-\frac{\omega_{\textrm{BD}}A\Phi_{\textrm{BD}}'^2}{2\Phi_{\textrm{BD}}^2}
-\frac{AB'\Phi_{\textrm{BD}}'}{2\Phi_{\textrm{BD}}B}\ ,
\end{align}
\begin{align}\label{26}
G_r^{\
r}=&\frac{(D-2)\omega_{\textrm{BD}}+D-1}{[(D-2)\omega_{\textrm{BD}}+D]\Phi_{\textrm{BD}}}
\times\nonumber\\ &\left[W-U-\frac{(D-2)W-DU
-2\Lambda\Phi_{\textrm{BD}}\frac{(D-2)\omega_{\textrm{BD}}+D}
{(D-2)\omega_{\textrm{BD}}+D-1}}{(D-2)\omega_{\textrm{BD}}+D-1}\right]
\nonumber\\
&+\frac{\omega_{\textrm{BD}}A\Phi_{\textrm{BD}}'^2}{2\Phi_{\textrm{BD}}^2}+
\frac{A\Phi_{\textrm{BD}}''}{\Phi_{\textrm{BD}}}+\frac{A'\Phi_{\textrm{BD}}'}
{2\Phi_{\textrm{BD}}}\ ,
\end{align}
and the equation of motion for the BD field:
\begin{align}\label{27}
A\left[\Phi_{\textrm{BD}}''+\left(\frac{D-2}{r}+\frac{A'}{2A}-\frac{B'}{2B}\right)
\Phi_{\textrm{BD}}'\right]=\nonumber\\ \frac{(D-2)W-DU
-2\Lambda\Phi_{\textrm{BD}}\frac{(D-2)\omega_{\textrm{BD}}+D}
{(D-2)\omega_{\textrm{BD}}+D-1} }{(D-2)\omega_{\textrm{BD}}+D}\ .
\end{align}
We numerically solve the coupled system of eqs. \ref{25}-\ref{27}.

We can find the total mass of the field configuration by the
relation:
\begin{equation}\label{28}
A(\rho)=1-\frac{2G_Dm_{\rho}}{\rho^{D-3}}-\frac{2\Lambda\rho^2}{(D-2)(D-1)}\
,
\end{equation}
where $m_{\rho}$ is straightforward connected to the total mass,
$M_{\rho}$, trapped within a sphere of radius $\rho$:
\begin{equation}\label{29}
M_{\rho}=\frac{D-2}{8\pi}\frac{2\pi^{(D-2)/2}}{\Gamma\left(\frac{D-1}{2}
\right)}m_{\rho}\ ,
\end{equation}
which, with our rescalings, gives for the total mass, $M$:
\begin{equation}\label{30}
M=(D-2)\frac{\pi^{(D-1)/2}}{\Gamma\left(\frac{D-1}{2}\right)}r^{D-3}\left[1-A(r)-
\frac{2\Lambda r^2}{(D-2)(D-1)} \right]\
,\hspace{1em}r\rightarrow\infty\ .
\end{equation}
In the case of a rotating star, one should embed General Relativity
in a more general framework, Metric-affine Gravity, where the
correct renormalized mass is (in 4 dimensions):
$$\frac{M}{1+\frac{\Lambda}{3}\frac{J^2}{M^2}}\ ,$$ with $J$ the
total angular momentum of the star, \cite{mag1,mag2}. In our case
the anti de Sitter mass is identical with the above renormalized
mass, because $J=0$. The mass of eqs. \ref{28}-\ref{30} is the
Schwarzschild mass, which corresponds to the ADM mass in the Jordan
frame. In the literature,
\cite{bosonstar-bransdicke3,bosonstar-bransdicke4}, apart from the
above mentioned mass, one also defines the ADM mass in the Einstein
frame, or tensor mass $M_T$, and the Keplerian mass, $M_K$. In the
framework of a scalar tensor gravitational theory, these masses are
obtained from the relations:
\begin{equation}\label{30a}
M_K=M+\Phi_1\hspace{1em}M_T=M+\frac{\Phi_1}{2}\ ,
\end{equation}
with
\begin{equation}\label{30b}
\Phi_1=\lim_{r\rightarrow\infty}(r^2\Phi'_{\textrm{BD}})\ .
\end{equation}
In our case we used the ADM mass in the Jordan frame for three
reasons: It comes naturally from the Schwarzschild mass, we can
deduce the Keplerian and tensor mass from eqs. \ref{30a} and
\ref{30b} and in our case is always larger from $M_K$ and $M_T$. For
the $\omega_{\textrm{BD}}=-1$ case, and according to the literature,
\cite{bosonstar-bransdicke3,bosonstar-bransdicke4},
$\Phi'_{\textrm{BD}}>0$ holds at infinity. For our case
($\omega_{\textrm{BD}}\geq5$), we found numerically that
$\Phi'_{\textrm{BD}}<0$ not only at infinity but also for the whole
region outside the soliton star. Because, in our case, $M>M_K,M_T$,
and because the energy of the same number of free particles is
larger than $M$, we find that the energy of the free particles is
always larger than $M_K$ and $M_T$, confirming the stability of the
soliton with respect to fission into free particles.

The Noether current that leads to the conserved particle number
is:
\begin{equation}\label{31}
j^{\mu}=\imath\sqrt{-g_D}
g^{\mu\nu}(\phi\partial_{\nu}\phi^{\ast}-\phi^{\ast}\partial_{\nu}\phi)\
,
\end{equation}
which gives a conserved Noether charge:
\begin{equation}\label{32}
N=\int_0^{\infty}d^{D-1}xj^t=
\frac{4\pi^{(D-1)/2}}{\Gamma\left(\frac{D-1}{2}\right)}\int_0^Rdr\omega\sigma^2r^{D-2}
\sqrt{\frac{B}{A}}\ .
\end{equation}

\begin{figure}
\centering
\includegraphics{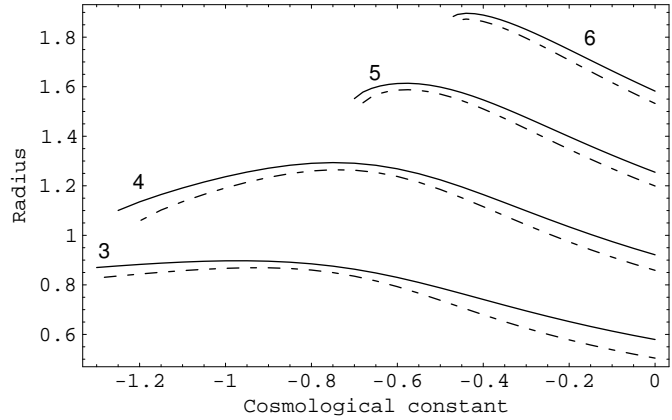}
\caption{The radius of a q-star as a function of the cosmological
constant. In figures \ref{figure6}-\ref{figure10} dashed lines
correspond to $\omega_{\textrm{BD}}=500$ and solid lines to
$\omega_{\textrm{BD}}=5$.} \label{figure6}
\end{figure}

\begin{figure}
\centering
\includegraphics{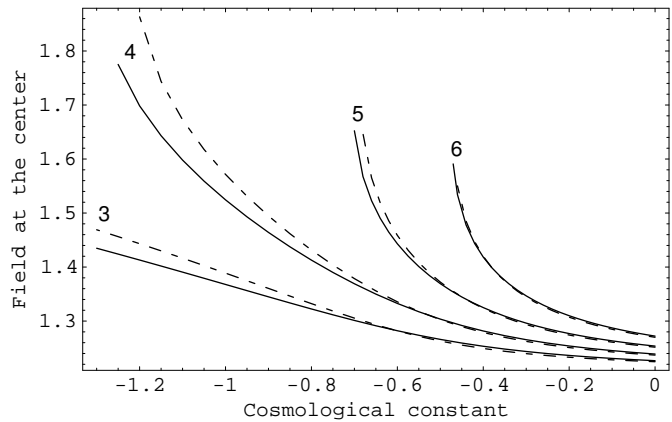}
\caption{The value of the matter scalar field at the center of a
q-star as a function of the cosmological constant.}
\label{figure7}
\end{figure}

\begin{figure}
\centering
\includegraphics{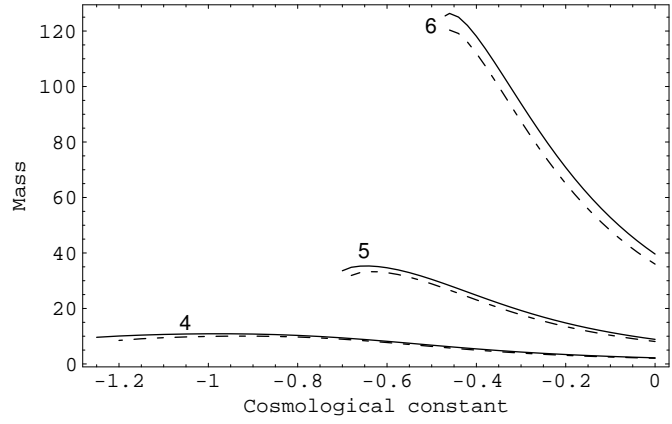}
\caption{The mass of a q-star as a function of the cosmological
constant.} \label{figure8}
\end{figure}

\begin{figure}
\centering
\includegraphics{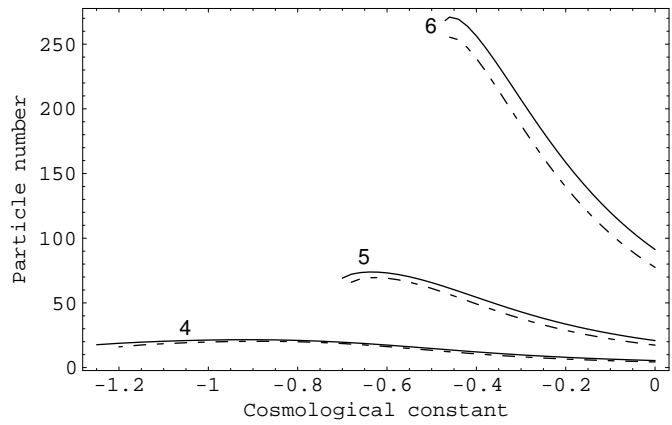}
\caption{The particle number of a q-star as a function of the
cosmological constant.} \label{figure9}
\end{figure}

\begin{figure}
\centering
\includegraphics{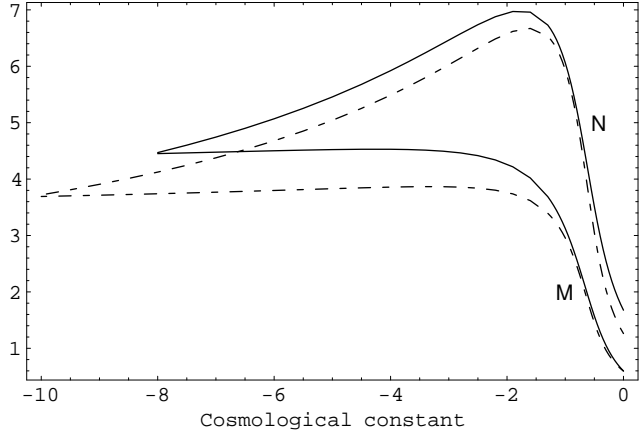}
\caption{The mass, $M$, and the particle number, $N$, of a q-star
in $2+1$ dimensions as a function of the cosmological constant.}
\label{figure10}
\end{figure}

\section{General scalar-tensor theory}

In the original BD gravitational theory $\omega_{\textrm{BD}}$ is
a constant. In a more general theory it may be regarded as a
function, usually of the BD scalar. We use one of the simple
functions, investigated in a cosmological framework,
\cite{scalartensor-cosmological,scalartensor-cosm-analytical},
namely:
\begin{equation}\label{s1}
(D-2)\omega_{\textrm{BD}}+D-1=\omega_0\phi_{\textrm{BD}}^n\ ,
\end{equation}
with $\omega_0$ and $n$ constants. The equation of motion for the
BD field is:
\begin{equation}\label{s2}
{\phi_{\textrm{BD}}}^{\hspace{1em};\lambda}_{;\lambda}=\frac{1}{(D-2)\omega_{\textrm{BD}}+D-1}
\left(8\pi T-\frac{d\omega_{\textrm{BD}}}{d\phi_{\textrm{BD}}}
{\phi_{\textrm{BD}}}^{,\rho}{\phi_{\textrm{BD}}}_{,\rho}\right)\ ,
\end{equation}
We set for simplicity $\Lambda=0$. We rescale:
\begin{equation}\label{s3}
\tilde{\omega}_0=\frac{(D-2)\omega_{\textrm{BD}}+D-1}{(D-2)\omega_{\textrm{BD}}+D}
G_D^n\omega_0\ ,
\end{equation}
and the other quantities as in eqs. \ref{15}-\ref{18}. We will
use for our calculations $n=1$. For $n=2,3,4$ the behavior of the
star parameters is similar. Dropping the tildes and the
$O(\epsilon)$ quantities we find for the Einstein equations:
\begin{align}\label{s4}
G_t^{\
t}=\frac{\omega_0}{\omega_0\Phi_{\textrm{BD}}+1}\left[-W-U-\frac{1}
{\omega_0\Phi_{\textrm{BD}}} \right.\times\nonumber\\ \left.
\left((D-2)W-DU-\frac{A\Phi_{\textrm{BD}}'^2}{D-2}
\frac{\omega_0\Phi_{\textrm{BD}}+1}{\Phi_{\textrm{BD}}}\right)\right]
\nonumber\\
-\frac{\omega_0\Phi_{\textrm{BD}}-D+1}{D-2}\frac{A\Phi_{\textrm{BD}}'^2}{2\Phi_{\textrm{BD}}^2}
-\frac{AB'\Phi_{\textrm{BD}}'}{2\Phi_{\textrm{BD}}B}\ ,
\end{align}
\begin{align}\label{s5}
G_r^{\
r}=\frac{\omega_0}{\omega_0\Phi_{\textrm{BD}}+1}\left[W-U-\frac{1}
{\omega_0\Phi_{\textrm{BD}}} \right.\times\nonumber\\ \left.
\left((D-2)W-DU-\frac{A\Phi_{\textrm{BD}}'^2}{D-2}
\frac{\omega_0\Phi_{\textrm{BD}}+1}{\Phi_{\textrm{BD}}}\right)\right]
\nonumber\\+\frac{\omega_0\Phi_{\textrm{BD}}-D+1}{D-2}\frac{A\Phi_{\textrm{BD}}'^2}
{2\Phi_{\textrm{BD}}^2}+\frac{A\Phi_{\textrm{BD}}''}{\Phi_{\textrm{BD}}}+
\frac{A'\Phi_{\textrm{BD}}'}{2\Phi_{\textrm{BD}}}\ .
\end{align}
The equation of motion for the BD field is:
\begin{align}\label{s6}
A\left[\Phi_{\textrm{BD}}''+\left(\frac{D-2}{r}+\frac{A'}{2A}-\frac{B'}{2B}\right)
\Phi_{\textrm{BD}}'\right]=\nonumber\\
\frac{1}{\omega_0\Phi_{\textrm{BD}}+1}
\left[(D-2)W-DU-\frac{A\Phi_{\textrm{BD}}'^2}{D-2}\frac{\omega_0\Phi_{\textrm{BD}}+1}
{\Phi_{\textrm{BD}}}\right]\ .
\end{align}

\begin{figure}
\centering
\includegraphics{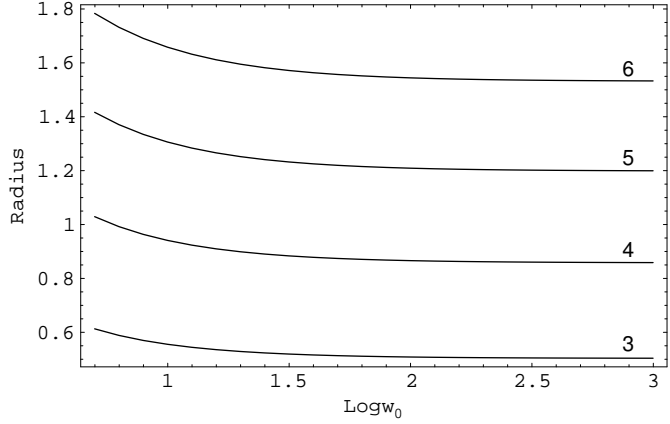}
\caption{The radius of a q-star as a function of $\omega_0$.}
\label{figure11}
\end{figure}

\begin{figure}
\centering
\includegraphics{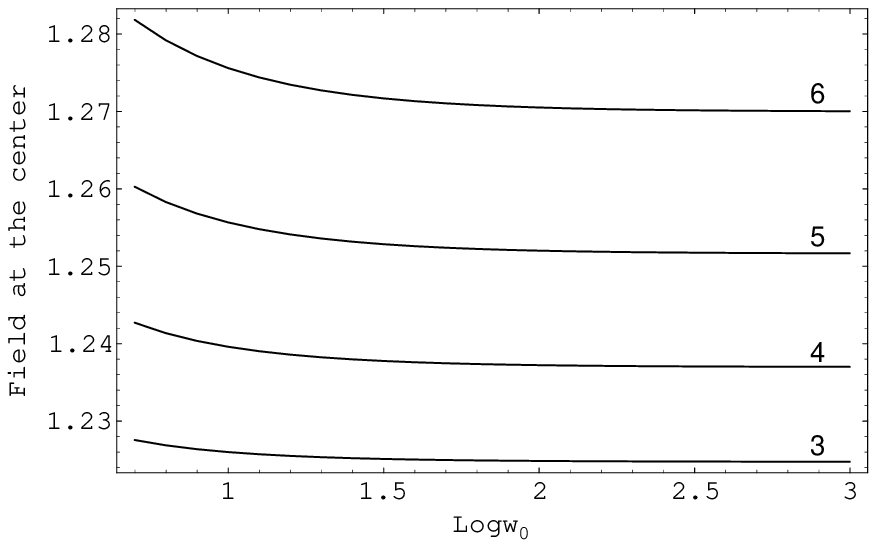}
\caption{The value of the matter scalar field at the center of
the q-star as a function of $\omega_0$.} \label{figure12}
\end{figure}

\begin{figure}
\centering
\includegraphics{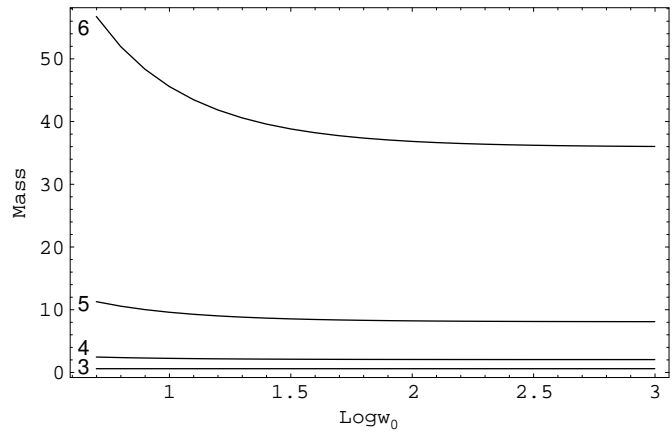}
\caption{The total energy of a q-star as a function of
$\omega_0$.} \label{figure13}
\end{figure}

\begin{figure}
\centering
\includegraphics{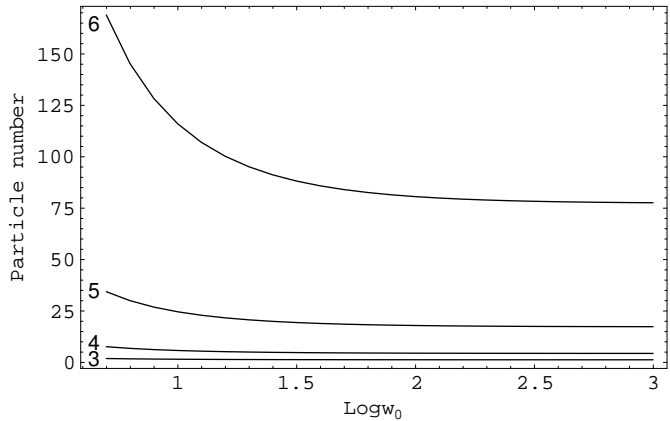}
\caption{The particle number of a q-star as a function of
$\omega_0$.} \label{figure14}
\end{figure}

We solve numerically the coupled system of eqs.
\ref{s4}-\ref{s6}. The mass and particle number of the star are
given by eqs. \ref{30}, \ref{32} respectively. We start from
$\omega_0=5$ up to $\omega_0=1000$, where the results of general
relativity are approximately reproduced.

\section{Conclusions}

In the present work we write the Brans-Dicke and a simple
scalar-tensor gravitational theory in $D$ dimensions and apply it in
a special case of stable, gravitating, scalar field configuration,
namely that of a q-star. The results of general relativity are
reproduced for $\omega_{\textrm{BD}},\omega_0\rightarrow\infty$. A
generally accepted, lower experimental limit is
$\omega_{\textrm{BD}}\simeq500$. Even for this value, the results
almost coincide with General Relativity and one can regard that the
results of General Relativity are practically obtained from our
figures when $\omega_{\textrm{BD}},\omega_0=1000$. We investigate
the phase space from $\omega_{\textrm{BD}},\omega_0=5$ up to 1000 so
as to gain a detailed picture of the behavior of the star
parameters, radius, value of the matter scalar field at the center
of the star, total mass and particle number. For
$\omega_{\textrm{BD}},\omega_0<500$ the star parameters are a few
per cent larger than in the case of General Relativity. These
differences are $D$-dependent. For example, for
$\omega_{\textrm{BD}}=5$ the particle number is 33, 24, 21, 18 per
cent larger from the corresponding results in General Relativity,
for $D=$3, 4, 5 and 6 dimensions respectively, the radius is 15, 7,
4 and 3 per cent larger respectively, and the total mass is $\sim$0,
7, 9 and 10 per cent larger. One can find similar differences for
the scalar-tensor theory.

The results of figures \ref{figure1}-\ref{figure5} and
\ref{figure11}-\ref{figure14} refer to zero cosmological constant.
Figure \ref{figure5} shows the ratio $M/N$ of the star as a function
of its particle number ($M$ is the ADM star mass and $N$ is the
particle number). As one can see, every field configuration is
stable with respect to fission into free particles. This feature is
characteristic for solitonic gravitating objects, which are stable
even in the absence of gravity. For example, a q-star with very
small particle number and mass corresponds to a (non-gravitating)
q-ball (in the thin-wall approximation). For small q-stars, one can
see from figure \ref{figure5} that $M/(mN)\rightarrow0.5$, with $m$
the mass of the free particles (here equal to unity with our
rescalings). This is the correct limiting value in the absence of
gravity, for the potential of \ref{19}, for which
$\omega_{\textrm{min}}\equiv(\sqrt{U/\sigma^2})_{\textrm{min}}=0.5$.
The picture is absolutely different for non-solitonic boson stars,
\cite{bifurcation1}, where gravity and not the scalar potential is
the main stabilizing factor. This means that for some region of the
phase space, gravity is not strong enough to make the binding energy
negative (equivalently, to make $M/(mN)<1$). In figures
\ref{figure6}-\ref{figure9} and for 4, 5 and 6 dimensions we start
from a zero value of the cosmological constant, gradually decrease
its value, and stop our calculations when $B(0)\rightarrow\infty$,
or $\sigma(0)\rightarrow\infty$, showing in this way the formation
of an anomaly at the center of the star. For $D=3$, when decreasing
the value of $\Lambda$ we find that both mass and particle number
increase up to their maximum value at approximately
$\Lambda\simeq-3$, and below this value, the energy is approximately
constant, when the particle number decreases slowly. Below
$\Lambda\simeq-10$ the energy of the free particles (identified with
the particle number for $m=1$) is less than the soliton energy,
which means that the decay into free particles is energetically
favorable. All the other field configurations depicted in our
figures are stable with respect to decay into free particles.

\vspace{1em}

\textbf{ACKNOWLEDGEMENTS}

\vspace{1em}

I wish to thank P. Manousselis and N. D. Tracas for helpful
discussions.

\end{document}